\documentclass{article}
\usepackage{graphicx} 
\usepackage{titlesec}
\usepackage{float}
\usepackage{url}
\usepackage{hyperref}
\usepackage{titlesec}
\usepackage{amsmath}
\usepackage{authblk}  
\makeatletter

\renewenvironment{abstract}{%
  \if@twocolumn
    \section*{\abstractname}%
  \else
    \small
    \begin{center}%
      {\bfseries \vspace{-2em}}
    \end{center}%
    \quotation
  \fi}
  {\if@twocolumn\else\endquotation\fi}
\makeatother

\title{Neuropsychology and Explainability of AI: A Distributional Approach to the Relationship Between Activation \& Similarity of Neural Categories in Synthetic Cognition
}
\author[1,2]{\textbf{Michael Pichat}}
\author[1,3]{\textbf{Enola Campoli}}
\author[1,4]{\textbf{William Pogrund}}
\author[1,5]{\textbf{Jourdan Wilson}}
\author[1,6]{\textbf{Michael Veillet-Guillem}}
\author[1,7]{\textbf{Anton Melkozerov}}
\author[1,8]{\textbf{Paloma Pichat}}
\author[1]{\textbf{Armanush Gasparian}}
\author[1,9]{\textbf{Samuel Demarchi}}
\author[1]{\textbf{Judicael Poumay}}

\affil[1]{Neocognition (Chrysippe R\&D) contact@neocognition.ai}
\affil[2]{University of Paris and Free Faculties of Philosophy and Psychology of Paris}
\affil[3]{Department of Cognitive Sciences and Department of Neuropsychology, University of Côte d’Azur}
\affil[4]{Department of Cognitive Sciences, University of Grenoble}
\affil[5]{Department of Linguistics, University of Paris Cité and University of California Los Angeles}
\affil[6]{Epitech Paris}
\affil[7]{Russian Academy of Sciences, FRC CSC RAS}
\affil[8]{Faculty of Medicine of Lyon East, University Lyon 1}
\affil[9]{Department of Psychology, University of Paris 8}

\begin{document}

\maketitle
\begin{abstract}
\noindent
We propose a neuropsychological approach to the explainability of artificial neural networks, which involves using concepts from human cognitive psychology as relevant heuristic references for developing synthetic explanatory frameworks that align with human modes of thought. The analogical concepts mobilized here, which are intended to create such an epistemological bridge, are those of categorization and similarity, as these notions are particularly suited to the categorical "nature" of the reconstructive information processing performed by artificial neural networks. Our study aims to reveal a unique process of synthetic cognition, that of the categorical convergence of highly activated tokens. We attempt to explain this process with the idea that the categorical segment created by a neuron is actually the result of a superposition of categorical sub-dimensions within its input vector space.
\end{abstract}

\section{Introduction}

Within an explainability framework, the neuropsychology of artificial intelligence focuses on studying synthetic neural cognitive mechanisms, considering them as new subjects of cognitive psychology research. The goal is to make artificial neural networks used in language models understandable by adapting concepts from human cognitive psychology to the interpretation of artificial neural cognition. In this context, the notion of categorization is particularly relevant because it plays a key role as a process of segmentation and reconstruction of informational data by the neural vectors of synthetic cognition.

Thus, in this study, the aim is to use the concept of categorization, as understood in human cognitive psychology (particularly in its relation to the notion of similarity), to apply it to the analysis of neural behavior and to infer certain synthetic cognitive processes underlying the observed behaviors.

\section{Categorical Explainability of Artificial Neural Networks}

\subsection{Epistemology and Utility of Synthetic Explainability}

Explainability involves describing the activity of an artificial neural network in terms that are understandable to humans (Du et al., 2019 ; Pichat, 2023, 2024a, 2024b). At the very least, it involves projecting the observable behavior of a neural network into an interpretative framework that allows for the assignment of meaning to this behavior, which is relevant to an observer, depending on their objectives. In our case, this framework is that of cognitive psychology, which involves using the categories of human cognitive thought (and, more specifically for us, the notion of categorization) as conceptual references to draw heuristic analogies between human and artificial cognitive behaviors. This must be done while avoiding the epistemological pitfalls of anthropomorphism (Nadeau, 1999), neo-behaviorism (Bloch et al., 2011), or the confusion between the observer and the observed system, against which cybernetics, systems theory, and enactive cognitive science warn us (Watzlawick, 1977, 1984; Varela, 1984, 1996).

The pragmatic function of explainability is twofold. First, it aims to inhibit potentially misleading or even dangerous responses from synthetic neural systems (Luo et al., 2024): errors, cognitive biases (Echterhoff, 2024) or cultural biases (Kheya, 2024), hallucinations (Kandpal et al., 2023; McKenna et al., 2023), undue focus on certain inputs (Du et al., 2023), etc. Second, it seeks to enhance the performance of language models (Bastings et al., 2022) by improving their coherence with human reasoning (Ma et al., 2023). In short, the goal is to develop cognitive alignment (Pichat, 2023, 2024a; Khamassi, 2024) of artificial cognition with human cognition, even if this sometimes involves a paradoxical and contradictory demand that artificial intelligence systems both exceed and respect human thought.

In this work, we focus on what can be epistemologically described as explainability with fine cognitive granularity, also known as mecanistic explainability. That is, an explainability that does not cognitively analyze the macroscopic level of a neural network’s final outputs based on its inputs; for example, its cognitive biases, cultural biases, or the effect of cognitive scaffolding in Bruner's (1990) sense of the "chain of thought" in prompt engineering (Zhen et al., 2024). Instead, it is a microscopic explainability where the unit of observation is the formal neuron (either alone or combined with others within intra- or inter-layer clusters). This low-granularity explanatory approach aims to delve into the internal cognitive system of the artificial neural network by creating elements of understanding regarding how categories of thought and concepts are more or less locally encoded and structured within the language model (Dalvi et al., 2019, 2022). The aim is, therefore, to interpret how categorical knowledge and even cognitive processes are developed and implemented by formal neurons (Fan et al., 2023).

\subsection{Objects of Study in Synthetic Explainability}

In the realm of fine-grained explainability, studies typically focus on either multilayer perceptron-type neurons (Bricken, 2023) or the attention heads of transformers (Clark et al., 2019). Regarding attention heads, the goal is to understand the type of categorical knowledge or cognitive processes encoded by the attention weights. Here, the interest may include, for example, the effects of attention on the analysis of data reliability in the initial and middle layers (Li et al., 2023), the attentional identification of syntactic categorical features such as indirect objects (Wang et al., 2022), or the mnemonic effect of attention weights (Geva et al., 2021).

In terms of the type of cognitive entities studied at the neuronal level, two classes of explainability studies can be distinguished: those related to cognitive contents (categories, concepts) and those focusing on cognitive processes (circuits). In the first category, the main subject of study is the relationship between neurons and specific conceptual categories (Sajjad et al., 2022; Foote et al., 2023). In the second category, studies examine, for example, the effect of connection weights between neurons on performing elementary logical operations (AND, OR, etc.) in certain neural branches (Voss et al., 2021); or subsets of neurons involved in decision-making (Antverg \& Belinkov, 2022).

In terms of neuronal scope, two types of investigations can be differentiated: those focusing on categorical encodings specific to isolated neurons or attention heads (Jaunet et al., 2021 ; Bills et al., 2023) and those with a broader scope, aiming to trace the effect of neural connections on the formation of specific circuits that are homogeneous in their cognitive activity (Olah et al., 2020; Anthropic, 2023; Bricken, 2023).

Finally, let's mention studies aimed at explaining static cognitive phenomena (the "captured" category or the cognitive process encapsulated within a neuron or an assembly of neurons) as opposed to those positioned from a dynamic perspective. The latter, for example, analyze the evolution of attentional information across attention heads and layers (Yeh et al., 2023), decompose token representations into \textit{n} intermediate vectors across layers (Modarressi et al., 2022; Yang et al., 2023); or track how, at the end of each layer, new embeddings are formed by combining the attention embedding and the perceptron embedding of a layer, in order to track the progressive internal processing of vector representations (Kobayashi et al., 2023).

\subsection{Examples of Low-Granularity Synthetic Categorical Explanations}

Various studies reveal or rather infer a variety of categories (linguistic, logical, positional, etc.) encoded within neurons and attention heads. We present some of these here, thus omitting works related to neurons involved in cognitive processes.

In the context of the classical experiment by Clark et al. (2019) on BERT, the authors highlight the converging linguistic functions of the attention heads from the same layers:
\begin{itemize}
    \item Identification of syntactic or morphological linguistic categories: direct objects of verbs, objects of nominal determiners, objects of propositions, objects of possessive pronouns, verbs modified by passive auxiliary verbs.
    \item Identification of linguistic categories related to coreference: antecedents of coreferential mentions (she/her, talks/negotiations).
    \item Identification of separator categories that enable text segmentation and delimitation: periods, the separator token ``SEP''.
    \item Identification of positional categories: next token, previous token.
\end{itemize}

In their fascinating study on GPT2-XL, Bills et al. (2023) showcase a series of unique neurons, highlighting for some their strong consideration of contextual elements:
\begin{itemize}
    \item Categorical neurons associated with any token from a specific lexical field. For example, a neuron activating for words describing movement involving feet (ran, walked, danced, kicked, hopped, stepping, tiptoed); or a neuron linked to things done correctly.
    \item Neurons associated with the category of phrases possessing a certain semantic valence, such as the ``simile'' neuron related to phrases involving certainty or confidence.
    \item Anomaly-detecting neurons, for example, the category of truncated or strange words.
    \item Neurons reacting to a specific token but only within a given linguistic context, thus forming a contextual category relative to a token. For instance, the ``hypothetical had'' neuron activating for the token ``had'' in a hypothetical context or in which things could have been different; or a neuron active for ``together'' but only when preceded by ``get''; or neurons becoming operational for specific words at the beginning of the text.
    \item Neurons detecting logical sequences such as the category of repetition of identical tokens or the category of breaking the logic of a sequence (1, 2, 3, 5).
    \item Anticipatory categorical neurons (clearly linked to the purpose for which the model was trained) becoming functional in a context corresponding to a likely next token, for example, when the next token is likely ``from''.
\end{itemize}

Studies also point to a geographical distribution of the specific type of neuronal categorical activity depending on the depth level of the layers they occupy. For example, the attention heads of the initial layers have broader attention compared to those more focused on tokens (Clark et al., 2023). Additionally, the initial layers are more responsive to categories of morphological elements at the word level, while later layers are more sensitive to syntactic categorical features related to sentences (passive/active voice, tense) and semantic categorical information (Jawahar et al., 2019).

\section[Synthetic Categorization and Human Categorization]{Synthetic Categorization and Human \\ Categorization}

\subsection[Main Categorical Characteristics of an Artificial Neural Network]{Main Categorical Characteristics of an Artificial \\ Neural Network}

An artificial neural network, particularly a language model, is organized into three components: input and output layers, which have perceptual and effector functions (Savioz et al., 2010), and intermediate layers, known as hidden layers. In the case of transformers, these intermediate layers include multilayer perceptron layers, which are the least studied (Garde et al., 2023) and are the focus of our current work, and attention layers.

Multilayer perceptron layers are characterized by an aggregation function and an activation function. The aggregation function, in the form ${\scriptstyle \sum (w_{i,j} x_{i,j}) + a}$ has specific parameters unique to each neuron and produces the categorical segmentation at the output (i.e., creates a new category) based on the input vector representation it receives. Initially, each weight $w_{i,j}$ in the aggregation vector indicates to what extent its associated categorical dimension j of the input semantic vector space should be selectively considered to generate the new output categorical dimension. In other words, each weight acts as an epistemological selector (Pichat, 2024b) that, from a cognitive perspective, performs an activity of selective categorical attention by governing the level of importance assigned to a given input categorical dimension. Secondly, all the products (attention weight x categorical dimension) are summed. This additive concatenation cognitively performs an activity of weighted epistemological fusion (Pichat, 2024) of the involved input categorical dimensions. The end result of this weighted linear combination is the creation of a new categorical segment (i.e., a new categorical dimension j', a new category), which is more abstract and relevant to the network's targeted activity. In a neurobiological analogy, weights are instantiated by synaptic efficacy, which is the amount a synapse releases its neurotransmitter into the synaptic cleft (Savioz et al., 2010).

The activation function (Sigmoid, Tanh, GeLU, ReLU, Leaky ReLU, Softmax, etc.), besides possibly normalizing the neuron outputs, introduces non-linearity into the neural system. From a cognitive perspective, this non-linearity increases categorical contrast (better signal-to-noise distinction) and thereby facilitates the convergence of neuron activation for certain categories (i.e., facilitates the construction of stable and distinguishing categories). It allows for a hierarchy in the distribution of categories constructed by the network (with more elementary categories in the early layers and progressively increasing complexity in subsequent layers) and a relative sparsity of the network. This sparsity, meaning the limited activation of neurons depending on the type of input, prevents overfitting (categories that are too specific, making generalization and adaptation to diversity impossible) and reduces the network's computational cost. The biological counterpart of the activation function is the transfer function (Savioz et al., 2010), of the type $\frac{1}{1 + \exp\left(-\left(G \cdot \text{activation\_net} + b\right)\right)}$, at the level of generating the action potential at the axon hillock and at the level of neuromodulation by dopamine and norepinephrine at the synapse (Servan-Schreiber, 1990).

Attention heads (Vaswani et al., 2017) rely on the fundamental mechanism of self-attention, which allows for the calculation of contextual attention weights for each categorical part of the input data using three learned weight matrices (queries Q, keys K, values V). The attention scores thus determined, by comparing each element with all other elements in a given informational sequence, enable informational selectivity (i.e., focusing attention on relevant categories of elements to the detriment of irrelevant ones) and contextual flexibility (i.e., adjusting the attention weight assigned to certain categorical elements based on the overall context and relationships between elements). With a fundamentally organizational purpose, each attention head captures or rather constructs a type of structural categorical relationship between different elements of an input sequence and then injects these "categorical meta-information" into the representations of the elements to be processed. These representations are thus informationally enriched by the categorical relationships that these elements maintain with other important elements in the sequence in question.

\subsection{Human Categorization}

In the realm of human thought, categorization is essential in various cognitive activities, such as classification, object identification, understanding, reasoning, problem-solving, memory, inference, prediction, and conceptualization (Sternberg, 2007; Roads et al., 2024).

Rosch (1975) proposes an approach to categorization based on the resemblance of an object to a prototype of the category, noting that individuals tend to state characteristic features rather than determinative properties (Rosch \& Mervis, 1975); the prototype being the most representative example of the category (Singh et al., 2020; Vogel et al., 2021). The exemplar theory of categorization (Medin \& Schaffer, 1978; Nosofsky, 1992; Nosofsky et al., 2022), as for it, suggests that objects are compared to typical examples stored in memory, with the most typical exemplar being the one that most closely resembles the known exemplars. Finally, the contextual or goal-directed approach to categorization (Barsalou, 1983; Glaser et al., 2020) emphasizes the purpose of the situation in defining a category rather than relying on a general logic or semantics.

The approaches to categorization by similarity, which are of particular interest in the context of this work, posit that an object is assimilated to a class based on its proximity to a representation of that class (Thibault, 1997; Jacob et al., 2021; Kaniuth et al., 2022; Roads et al., 2021, 2024). The prototype and exemplar theories mentioned earlier emphasize similarity as the basis for categorization (Ayeldeen et al., 2015 ; Sanborn et al., 2021; Roads et al., 2024). However, critics of similarity-based categorization argue that the choice of criteria for judging similarity is arbitrary and may not align with the foundations of categorical attribution (Love, 2002; Kalyan et al., 2012; Reppa et al., 2013; Poth, 2023). Reasoning by similarity is thus considered too ambiguous to be functional (Wixted, 2018). Proponents of similarity in categorization (Bobadilla et al., 2020; Hebart et al., 2020) counter-argue that these criticisms are based on two epistemological errors: (i) a realist error that assumes categorization must capture a predefined reality, and (ii) a rationalist error that imposes a normative logic on categorization, a logic that every individual is expected to understand.

\section{Problem Statement}

As we have mentioned, the goal of synthetic explainability is to make the operations of an artificial neural network accessible to human understanding (Du et al., 2019). This requires converting the observable behavior of neural networks into an interpretative framework containing explanatory elements compatible with the cognitive frameworks of human thought. In this context, human cognitive psychology emerges as a relevant heuristic framework for developing explanatory analogies of synthetic cognition. More specifically, the cognitive psychology of categorization is particularly pertinent because synthetic information processing significantly involves segmentation and categorical analysis behavior (Pichat et al., 2024). Indeed, numerous studies (Jawahar et al., 2019; Clark et al., 2019; Bills et al., 2023; Clark et al., 2023) highlight artificial cognition in language models as relying largely on a dynamic extraction of broad linguistic categorical invariants.

In the context of this study, as previously mentioned, we focus on an epistemological explainability with fine cognitive granularity (Pichat, 2024). In other words, we examine a microscopic explainability where the unit of observation is the formal neuron. This low-granularity interpretative approach aims to directly penetrate the "black box" system that constitutes an artificial neural network, creating elements of understanding about how categories of thought and concepts are locally encoded and structured within a language model (Dalvi et al., 2019, 2022). The objective is to interpret how categorical knowledge is constructed and utilized by the fundamental elements of the networks, namely the neurons themselves (Fan et al., 2023). Concerning the specific question of utilization, we focus here on the particular issue of the relationship between activation and categorical similarity.

Indeed, in connection with the problem we have raised in the field of human cognition regarding the relationship between categorization and similarity, we have, in a previous study (Pichat et al., 2024), transposed this relational heuristic question to the domain of artificial cognition by posing the following inquiry: Is the degree of membership (operationalized in terms of activation level) of tokens (received by a neuron in the form of embeddings) to the category associated with that neuron related to their level of similarity (operationalized in terms of cosine similarity)? In other words, are the intensity of categorical membership and the intensity of similarity of tokens, as evaluated by a neuron, two facets of the same phenomenon? This question, largely unexplored to date in the field of artificial systems explainability (Fan et al., 2023; Luo et al., 2024; Zhao et al., 2024), seemed particularly relevant to examine.

In our previous study, we demonstrated the compatibility of our results with two formulated hypotheses: (i) a categorical discontinuity of successive core (i.e. high mean level of activation) tokens in terms of activation level (suggesting particularly low cosine similarities between these core tokens) and (ii) a categorical heterogeneity of core tokens with similar activation levels (these core tokens are not the closest in terms of cosine similarity). These two hypotheses complement each other and both address the overall relationship between activation proximity and cosine similarity. However, they are positioned within a static approach, aiming to investigate the relationship between activation proximity and cosine similarity independently of the activation level of the tokens involved. However, other results, not yet explored during this initial research, prompt us to continue this investigation from a distributional perspective, questioning a possible evolution of the relationship between categorical membership proximity (i.e., activation proximity) and categorical proximity (i.e., cosine similarity) depending on the activation levels of the tokens involved. This is the work we undertake in the present research.

\section{Methodology}

\subsection{Methodologies of Synthetic Explainability}

To provide methodological context for our study, we present here a brief, non-exhaustive overview of technical approaches, whether at low or high granularity, aimed at inferring the cognitive contents or processes encapsulated within formal neurons and their assemblies (in layers, clusters, or the global network). These methods are not mutually exclusive, allowing for a certain degree of overlap.

High-granularity techniques, as previously mentioned, are grounded in the input/output contrast and aim to study the relationship between initial information and final outputs of a language model. In this context, gradient-based methods seek to measure the importance of each input by analyzing the partial derivatives of the output with respect to each input dimension (Enguehard, 2023). Input characteristics can be measured in terms of features (Danilevsky et al., 2020), token importance scores (Enguehard, 2023), or attention weights (Barkan et al., 2021). Similarly, example-based approaches aim to understand the extent to which the output changes with different inputs. This is done by showing how the network outputs are impacted by small changes in input (Wang et al., 2022) or by alterations (deletion, negation, mixing, masking) of inputs (Atanasova et al., 2020; Wu et al., 2020; Treviso et al., 2023). Lastly, we should mention works that perform a conceptual mapping of inputs and then measure the contribution of these concepts to the observed outputs (Captum, 2022).

Fine-granularity methodologies, as discussed earlier, do not take the final output of the language model as their output but rather its intermediate outputs or states at the level of neurons or clusters or layers of neurons. In this context, some methods aim to linearly decompose the relevance score of a neuron in a given layer based on its inputs (neurons, attention heads, or tokens) in the previous layer (Voita et al., 2021). Other methods focus on linearizing activation functions to facilitate neural interpretation (Wang et al., 2022). Still other methods, based on the model's vocabulary, seek to identify the knowledge encapsulated by projecting the connection weights and intermediate representations into the model's vocabulary space via a de-embedding matrix (Dar et al., 2023; Geva et al., 2023). Finally, we should mention approaches based on neuronal activation statistics in response to corpora (Durrani et al., 2022; Wang et al., 2022; Dai et al., 2022; Bills et al., 2023; Mousi et al., 2023). It is within the specific context of these latter approaches that our present study is situated.

\subsection{The Explainability Study as the Source of Our Data}

Our data is derived from the intriguing study by Bills et al. (2023). Based on the hypothesis that a neuron activates specifically for a property to be determined, which may include context, the authors conducted an extensive study to interpret the categorical semantics of all neurons in GPT-2XL.

Methodologically, the OpenAI researchers proceed as follows. They subject GPT-2XL (the "subject" model) to a large series of 64-token sequences, randomly extracted from the internet data with which the model was trained. For each token, its activation values for all neurons across all layers are recorded. A more advanced model, GPT-4, the "explainer" model, is then employed to automatically identify the elements to which each neuron responds (i.e., to generate the "explanation") based on an instructional and example prompt (few-shot learning) operating solely on the five text sequences with maximum activation (i.e., containing at least one token with maximum activation), determined by the quantile of maximum activations. GPT-4 is subsequently used as a "simulator" model: based on a prompt providing the "explanation" of each neuron, the simulator must predict the activation level of each token for the same 64-token sequences. Finally, for each neuron, its actual and predicted activations for each token are compared to calculate an explanation score intended to measure the quality of the generated interpretation.

The main results of the study are as follows. Concerning the tokens used for neural explanation: focusing on the top 5 activations of the neurons is interpreted by the authors as the most effective in terms of prediction score, and increasing the number of tokens does not significantly improve the prediction score; the inclusion of tokens with lower activations decreases the prediction score. Regarding the quantitative dimension of the explanation scores: the average explanation score is low at 0.15 (only 1,000 neurons out of 307,200 have a score above 0.8), this score decreases with the depth of the layers, and the explanation scores increase with the model explainer's complexity and sparsity. Regarding the qualitative dimension of the explanation scores: both GPT-4 and human explanations show low prediction scores; the automatically generated explanations are too broad (descriptive hypernyms are too general where hyponyms would be more specific to the precise data involved).

In the context of our current work, we revisit the activation data, neuron by neuron, obtained for the wide range of tokens used in the study conducted by Bills et al. (idem). We reuse this data for other purposes, aiming to study, as mentioned, the relationship between activation and similarity at the level of synthetic neuronal categorical cognition.

\subsection{Selection and Interpretation of Data}

We briefly present the methodological choices made in this current study, in continuity with those made during our previous study (Pichat et al., 2024) concerning the relationship between categorical membership (activation) and categorical proximity (similarity), which this current work continues.

To simplify our study, we limited our analysis to the first two layers of GPT-2XL (layers 0 and 1) and the 6,400 neurons in each layer. For each neuron among these 12,800 neurons in total, we chose to consider its 100 most highly activated tokens on average, along with their respective activation values. This approach differs from that of Bills et al. (2023), which focuses only on hyperactivated tokens. We consider this method partially limited (even though it allows for valuable effects, as indicated by the authors) because it does not capture the variability of tokens for which a neuron activates. We prefer a more comprehensive view of the category of tokens to which a neuron responds.

We consider the average activation level of a token in a neuron as a good measure of the categorical membership of that token to the involved neural category. Indeed, the average activation of the most activated tokens (i.e. core tokens) seems to represent well the extent to which these tokens are part of the extension of a category. This is in line with the hypothesis of Bills et al. (idem) that a neuron activates for a specific property.

The cosine similarity between two tokens also seems to be a good measure of categorical similarity between items. This aligns with Thibault (1997), who defines similarity based on a method of calculating the distance between compared categorical dimensions. Cosine similarity, commonly used in NLP to measure semantic proximity (Ham, 2023), fits well with this definition.

We chose to measure cosine similarity within the embedding base of GPT-2XL to avoid the methodological limitations mentioned by Bills et al. (2023) and Bricken (2023), which involve matching synthetic cognitive systems based on different embedding systems. For comparison and verification, we also used three other freely available embedding bases: Alibaba-NLP/gte-large-en-v1.5, Mixedbread-ai/mxbai-embed-large-v1, and WhereIsAI/UAE-Large-V1.

\section{Results}
\subsection[Evolution of the Relationship Between Cosine Proximity and Activation]{Evolution of the Relationship Between Cosine \\ Proximity and Activation}

To recap, we study the relationship between activation proximity and cosine proximity for successive core tokens of each neuron, that is, the relationship between the proximity of categorical membership levels between two tokens and the categorical proximity between these two tokens. Specifically, we focus on the dynamics of this distribution according to the levels of activation. In other words, we investigate whether there is a potential evolution, across activation segments, in the relationship between cosine proximity and the activation levels of successive core tokens. Graphs 1 (layer 0) and 2 (layer 1) show the average distribution of cosine similarities of pairs of successive core tokens according to the activation rank of the first token of each pair. We can clearly observe an evolution that is initially very slow (ranks 0 to 40), then faster (ranks 40 to 80), and finally appears exponential (from rank 80 onwards) in categorical proximity as a function of activation proximity.

Several points can be noted: (i) this evolution seems invariant across the four embedding bases (although the growth dynamic is more pronounced for GPT-2XL embeddings, as they are more discriminative, as mentioned in Pichat et al., 2024), (ii) the segment of exponential growth appears even more marked for layer 1 compared to layer 0 (is there an acceleration to be noted with increasing layer depth?), (iii) from rank 80 for layer 0 and rank 40 for layer 1, the average cosine similarities per rank become higher than the overall averages (respectively .38 and .42 according to the GPT-2XL embeddings, see tables 1 and 2), (iv) despite this, the growth in cosine proximity as a function of activation proximity is still bounded by not very high maxima of cosine similarity (respectively .48 and .55 with GPT-2XL embeddings).

\begin{figure}[ht]
  \centering
  \includegraphics[width=1\textwidth]{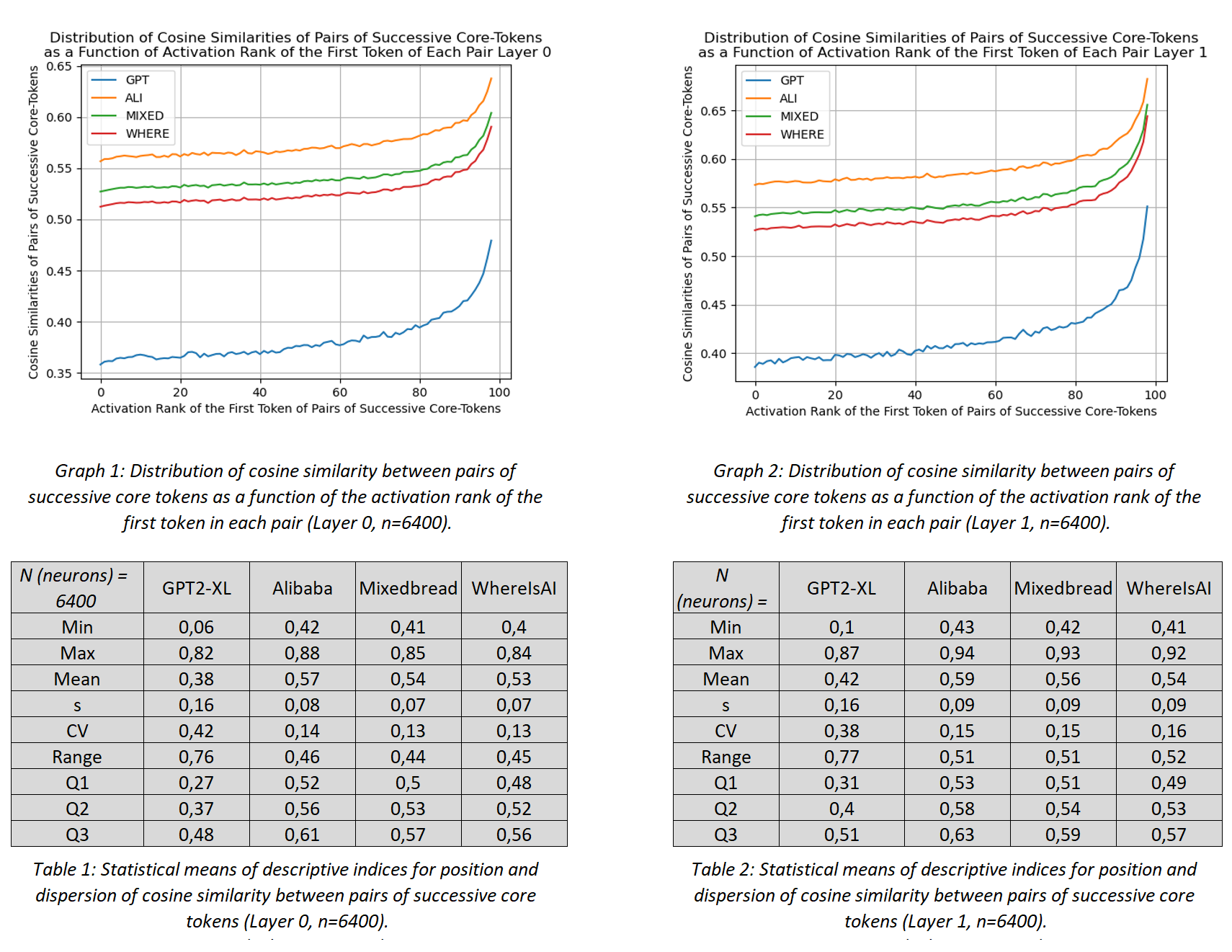}
\end{figure}

This initial exploratory and descriptive view of the positive monotonic evolution of the cosine similarity distribution as a function of the activation level leads us to formally investigate the hypothesis of the following synthetic cognition phenomenon: the categorical convergence of pairs of successive core tokens (in terms of their activation level) as these activation levels increase. In other words, the hypothesis posits that as the activation levels of successive core tokens (i.e., tokens that are close in terms of activation) increase, the categorical variability of these core tokens decreases (i.e., categorical proximity increases). We will test this hypothesis below using different operationalizations.

\subsection[Categorical Convergence and Extreme Values of Cosine Similarity]{Categorical Convergence and Extreme Values of \\ Cosine Similarity}

A first operationalization to test our hypothesis of categorical convergence of pairs of successive core tokens as activation levels increase can be based on the study of the distribution of lower extreme values of cosine similarity. Indeed, according to this hypothesis, the number of cosine similarity \textit{minima} should decrease as activation values increase.

Tables 3 (layer 0) and 4 (layer 1) present the distribution of the number of lower outliers of cosine similarity for pairs of successive core tokens based on the activation quartiles of the first token of these pairs (outliers derived from GPT-2XL embeddings with interquartile range). For layer 0, there is an overrepresentation (+.35) of weighted differences related to the first activation quartile segment, gradually decreasing to an underrepresentation (-.46) of weighted differences related to the last segment. Similarly, for layer 1, we find an overrepresentation of +.33, gradually decreasing to an underrepresentation of -.36. In both cases, the inferential $\chi^2$ adjustment analysis, with a theoretical equidistribution of 25\% consistent with our quartile segmentation, highlights this decrease \(p(\chi^2) < 0.05\).

\begin{figure}[ht]
  \centering
  \includegraphics[width=1\textwidth]{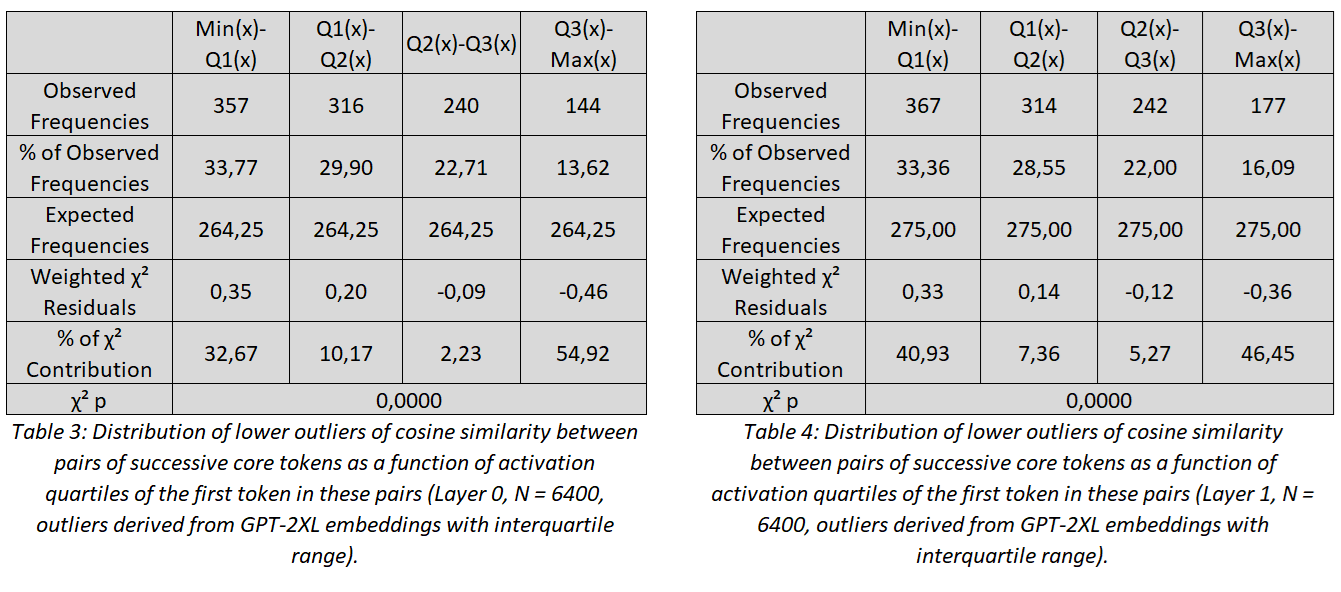}
\end{figure}

Following the same logic, let us now consider the low cosine values, which we define, neuron by neuron, as those below the threshold of the neuron's minimum cosine increased by 10\% of its range. Tables 5 (layer 0) and 6 (layer 1) demonstrate the same trend of progressive decrease in the number of low cosine values across activation quartiles; this trend remains significant \(p(\chi^2) < 0.05\).

\begin{figure}[H]
  \centering
  \includegraphics[width=1\textwidth]{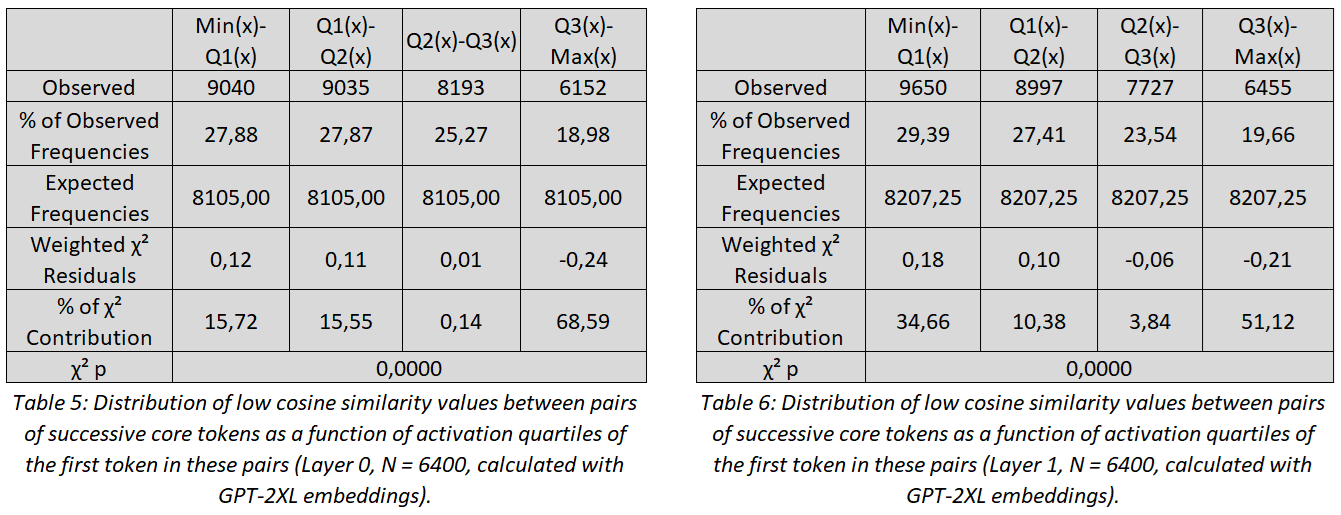}
\end{figure}

These two sets of results highlight the fact that the number of low extreme cosine similarity values decreases as activation values increase. They are thus compatible with our hypothesis of categorical convergence, which posits that as the activation levels of successive core tokens increase, the categorical variability of these core tokens decreases.

\subsection{Categorical Convergence and Positivity of Relational Monotonicity}

A second relevant operationalization to test our hypothesis of categorical convergence of pairs of successive core tokens as activation levels increase involves a more functional approach (in the mathematical sense) by studying the monotonicity and direction of the relationship between cosine similarity and categorical activation value. In line with our hypothesis, we should observe a monotonic and positive relationship (an increasing function) between these two variables.

A first approach in this context is to conduct a linear regression study. Tables 7 (layer 0) and 8 (layer 1) present the main terms of the linear regression of cosine similarity for pairs of successive core tokens as a function of the activation of the first token in each pair. At first glance, there is a variable and moderate compatibility with the condition for applying regression, which is the normality of residuals (the percentages of associated inferential tests with $p > .05$ vary from 62\% to 99\% for layer 0 and from 53\% to 98\% for layer 1). As a result, these regression data are not very reliable (see also graphs 3 and 4, which seem to support the same caution). A linear model is not well-suited to our data, with only 34\% of Fisher tests being significant for layer 0 and 45\% for layer 1. However, when changing the scale of statistical units (i.e., moving from tokens to neurons), several interesting results emerge. Firstly, the average slope coefficient \textit{a} is positive, albeit small (.06 for layer 0 and .04 for layer 1 with GPT-2XL embeddings), and relatively stable across the four embedding systems. Secondly, the percentage of neurons with such a positive slope coefficient is extremely high (80\% for layer 0 and 85\% for layer 1 with GPT-2XL embeddings), and this is fully consistent across the other three embedding systems; this trend is highly significant at the inferential level based on a $\chi^2$ goodness-of-fit test using a theoretical dichotomous equidistribution \(p(\chi^2) < 0.05\) for both layers). Given our caution regarding applicability, this last result could be compatible with our hypothesis, even though, again, the choice of a linear approach does not seem well-suited here. For illustration, graph 5 provides an example of a neuronal linear regression, showing the slight positive slope connecting the two variables in line with our hypothesis (see the annexes for an illustration using representative neurons from layers 0 and 1).

\begin{figure}[htbp]
  \centering
  \includegraphics[width=1\textwidth]{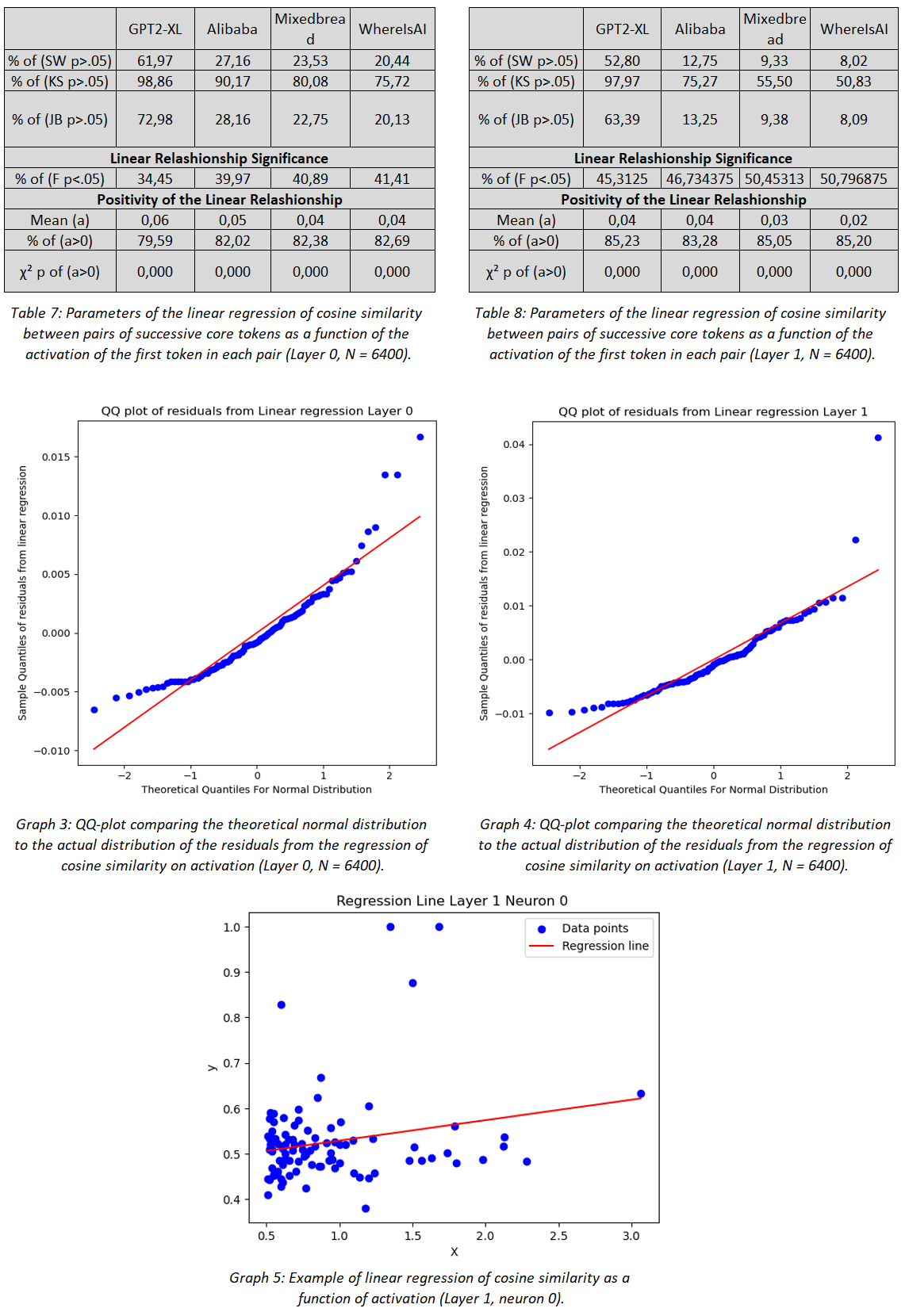}
\end{figure}

 We complement our initial methodological approach above with a second, non-parametric method (thus removing the conditions of normality) that may be more suitable since it is ordinal, using Spearman's $\rho$. We find the same type of results as before in Tables 9 and 10, which relate to the parameters of the ordinal relationship of cosine similarity for pairs of successive core tokens based on the activation of the first token in each pair. A relatively low percentage of cases where the effect of ordinal correlation is significant (33\% for layer 0 and 40\% for layer 1) again shows an insufficient relevance of this new modeling method chosen to account for our hypothesized effect. However, the average $\rho$ values are higher than their previous linear slope coefficient counterparts (.10 for layer 0 and .13 for layer 1, with GPT-2XL embeddings, stable values for other embeddings), indicating a slightly better fit of our ordinal modeling and still showing a positive relationship between our two variables. Again, a significant percentage of neurons is associated with a positive ordinal correlation coefficient (75\% for layer 0 and 81\% for layer 1), a result that is again significant (p($\chi^2) < 0.05$ for both layers).

\begin{figure}[ht]
  \centering
  \includegraphics[width=1\textwidth]{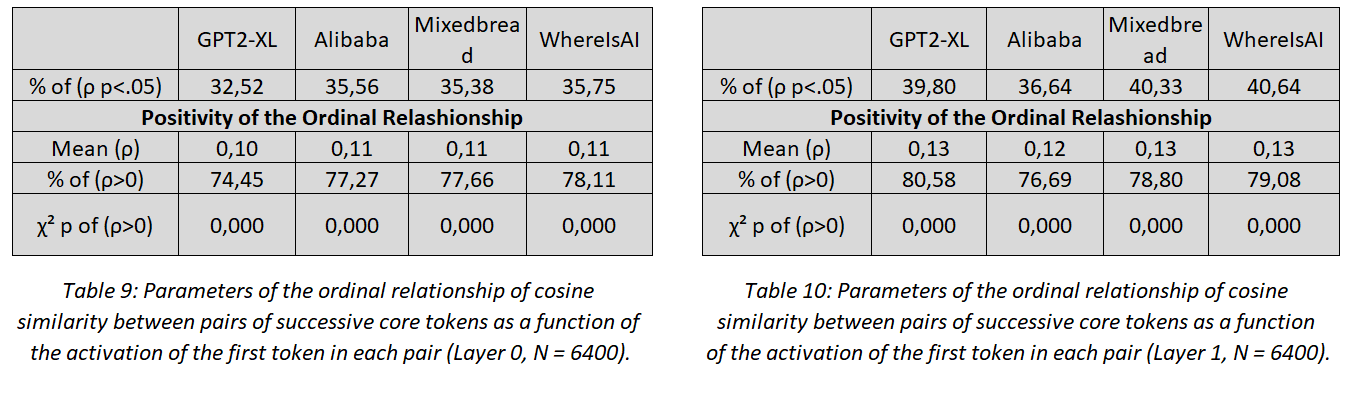}
\end{figure}

 The results of the two complementary methodological approaches we presented above are partially compatible with our hypothesis of categorical convergence for pairs of successive core tokens as activation levels increase.

Let us note that a renewed look at graphs 1 and 2, which present the distribution of cosine similarities between pairs of successive core tokens as a function of the activation rank of the first token in each pair, shows that a more suitable model is actually of the form \( y = a + \exp\left(\frac{x}{b} + c\right) \) (where \( y \) represents cosine similarity and \( x \) represents the activation value). The positive parameter \( b \) expresses the exponential growth rate of cosine similarity (the smaller \( b \) is, the faster this growth rate), and the negative parameter \( c \) indicates the activation zone corresponding to the onset of the exponential growth break of cosine similarity; parameter \( a \) denotes the y-intercept, which is the minimal cosine similarity associated with the minimum activation. Thus, an exponential regression analysis is far better suited than our linear or ordinal approaches to account for a positive monotonicity in the cosine/activation relationship, under the conditions we described in graphs 1 and 2, showing a very slow initial evolution of cosine with activation, then faster and finally appearing exponential.

\subsection{Categorical Convergence and Contrasting Activation Averages}

A third and final relevant operationalization to test our hypothesis of categorical convergence for successive core tokens as a function of activation involves a contrast-based approach, comparing the categorical proximity of pairs of successive core tokens between groups of tokens that are extremized in terms of their activation level. According to our hypothesis, for each neuron, the average cosine similarity of pairs of successive core tokens should be higher for pairs with high activations compared to those with low activations.

We use a non-parametric approach, the Wilcoxon-Mann-Whitney test, given the relative normality of our data and the small sample sizes of the groups we will use here. Tables 11 (layer 0) and 12 (layer 1) contain the main results concerning the mean differences in cosine similarities for the 21 pairs of successive core tokens with the lowest/highest activation ranks of the first token in each pair. We can observe a systematic global mean difference for the four embeddings, although the differences obtained with GPT-2XL embeddings are more pronounced due to this embedding model being more differentiating (cos$_{\text{min}}$=.364 / cos$_{\text{max}}$=.417 on average for layer 0; cos$_{\text{min}}$=.393 / cos$_{\text{max}}$=.459 for layer 1). However, there is a significance of this difference (\(p(\text{MW}) < .05\)) for only 30\% of the neurons in layer 0 (associated with a mean effect size that is effectively negative and non-negligible at -.17) and 37\% in layer 1 (associated with an effect size that is effectively negative and non-negligible at -.21); also, a stronger contrast for layer 1 compared to layer 0 raises the question of a potential intensification of this divide as layers deepen. Even if this contrast is not significant, it is worth noting that, in a large majority of cases for the four embeddings, we observe a higher average cosine similarity in groups of core tokens with high activations compared to low activations (76\% of neurons in layer 0 and 82\% in layer 1). In all cases (both layers and four embeddings), there is a significant effect of these percentages based on a 
$\chi^2$ goodness-of-fit test using a theoretical dichotomous equidistribution \(p(\chi^2) < 0.05\). These results are again compatible with our hypothesis of categorical convergence for pairs of successive core tokens as activation levels increase. It should be noted that a much stronger contrast would likely have been found neuron by neuron if we had used groups of tokens that were more extremized in terms of their activation, for example, by taking as a class of low activation tokens those that are not core tokens, that is, the 100 tokens most activated on average per neuron.

\begin{figure}[ht]
  \centering
  \includegraphics[width=1\textwidth]{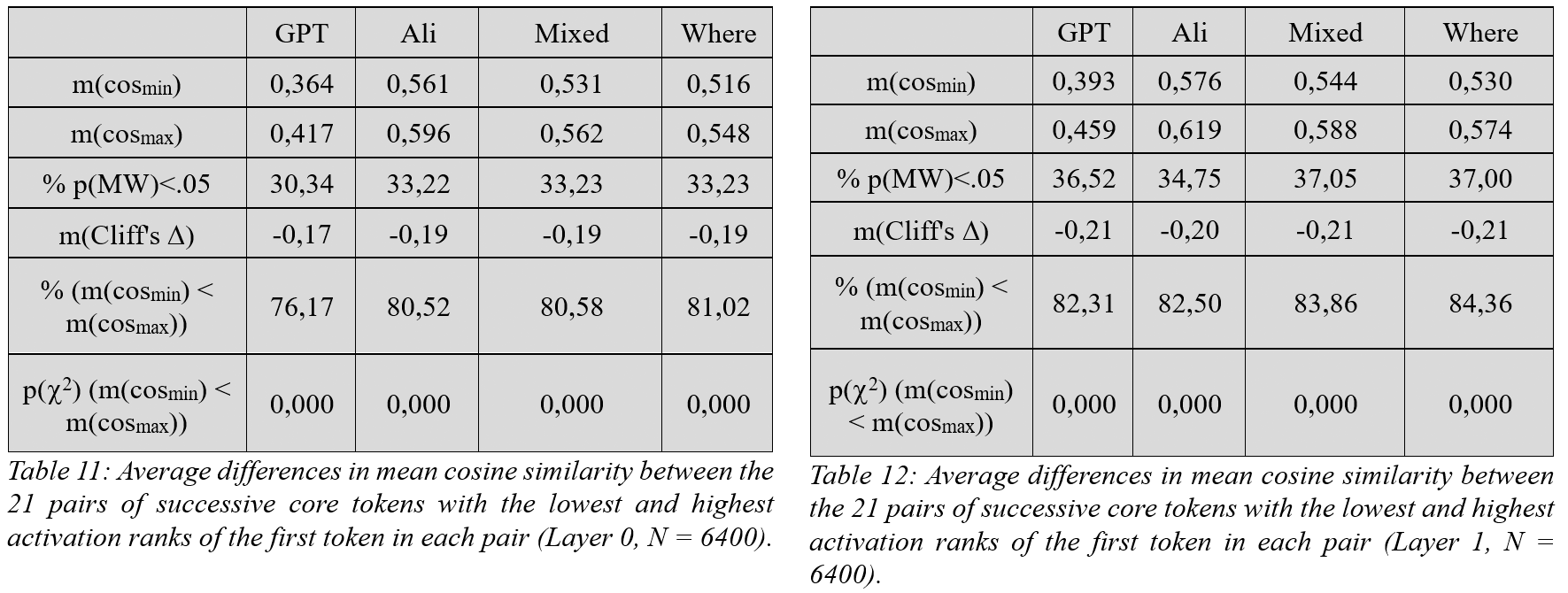}
\end{figure}

 At the conclusion of our three complementary operationalizations for testing our hypothesis of categorical convergence—suggesting that as the activation levels of successive core tokens increase, their categorical variability decreases (i.e., categorical proximity increases)—a significant portion of our statistical analyses appears to support this hypothesis.

\section{Discussion}
\subsection[Categorical Convergence for High Activations as Indicative of Co-Activation of Sub-Dimensional Categories within Synthetic Categories]{Categorical Convergence for High Activations as \\ Indicative of Co-Activation of Sub-Dimensional \\ Categories within Synthetic Categories}

We explored, regarding successive core tokens, the possible existence of a changing relationship between cosine proximity and activation. Our results in this area suggest a notion of categorical convergence for pairs of successive core tokens based on activation; in other words, as the activation levels of successive core tokens (i.e., those close in activation level) increase, their categorical variability decreases (i.e., categorical proximity increases). How should we interpret this result? In what follows, we will utilize various partial explanatory frameworks before attempting to coordinate them to answer this question.

In their work on the relationship between categorization and similarity in human cognitive psychology, Roads et al. (2024) emphasize geometric mental representations, meaning those for which an object's representation (to be categorized or compared for similarity with another) is expressed through multidimensional coordinates in a multi-axis space. This mode of representation aligns with the sub-dimensional categories related to a given neuron that we will discuss below. In this type of representational approach, Thibault (1997) states that studies linking categorization and similarity operate on the principle that an object is assimilated into a class by estimating its proximity to what represents that class; this is based on (i) a space of dimensions considered relevant for comparison and (ii) a method for calculating the distance between compared elements.

As mentioned earlier, those advocating a separation between categorization and similarity (Murphy and Medin, 1985; Rips, 1989; Barsalou, 1991; Medin et al., 1993; Love, 2002; Kalyan et al., 2012; Reppa et al., 2013; Poth, 2023) argue against the subjectivity, instability, and versatility of the criteria chosen in the moment to base a similarity judgment, criteria that represent only a singular and subjective choice among other possible dimensions. These objections highlight that reasoning by similarity is ambiguous, i.e., not sufficiently constrained (Goodman, 1972; Wixted, 2018).

However, as discussed, Goldstone (1994) counter-argues that similarity is not always unstable, and Thibault (1997) denounces a psychological essentialism in these critics, reproaching them that their argument against similarity's validity assumes that the criteria for categorical segmentation should have an intrinsic and stable value, based on the idea of copying an ontologically defined reality. Furthermore, Hampton (1997), taking a stance from fuzzy logic, responds that the criteria for categorization cannot be thought of as univocal and impermeable, that is, as pertaining to classical logic. This is in line with the positions of various defenders of similarity in categorization (Bobadilla et al., 2020; Hebart et al., 2020).

These different contradictions seem, in the end, partially aligned with our observation of the convergence of categorical similarity for pairs of successive core tokens based on activation. Indeed, this convergence beyond a certain activation threshold expresses the following discrepancy: there is no or a low relationship between categorical similarity and categorical proximity for "low" levels of activation, while a link between these two phenomena appears for higher activations. And this, precisely due to the types of arguments and counter-arguments summarized above, which we will now transpose to synthetic cognition, deliberately maintaining their antagonism.

As mentioned in our previous study (Pichat et al., 2024), successive tokens at low activation levels are associated with low cosine similarity values between them for a given neuron because they correspond to a variability, a diversity of sub-dimensional categories associated with that neuron; a given token pertains to a certain sub-dimensional category, while its neighbor in terms of activation pertains to another sub-dimensional category, resulting in a low cosine similarity reflecting categorical heterogeneity or disparity (at least, from our human perspective). This aligns with the arguments mentioned earlier against a link between categorization and similarity that invoke the versatility of dimensions used to assess similarity. In other words, a neuron encodes a complex, multi-dimensional category; a category that is not homogeneous but polysemic (Fan et al., 2023; Bills et al., 2023). For a given neuron, this polysemy makes us perceive this category as an alien concept (Bills et al., 2023) (which it effectively is from our human cognition) because it results from a superposition of different sub-dimensional categories generated by the intermediate vector base of this neuron (a base we conceive differently from Bricken et al. (2023) but rather, for a given neuron in layer n, in terms of the output categorical dimensions of its precursor neurons in  layer $n-1$).

Meanwhile, successive tokens with high activation levels are associated with higher cosine similarity values between them. This is because, by the mathematical construction of the aggregation function, they have a high activation level precisely because they jointly, simultaneously belong to several co-activated sub-dimensional categories (derived from their precursor neurons). Co-activations of different sub-dimensional categories can only occur insofar as, in the specific cases of the highly activated tokens involved here, these different sub-dimensional categories happen to be categorically, semantically convergent for the tokens involved; in other words, in particular situations where the tokens in question are at the semantic intersection of these diverse sub-dimensional categories. These categorical intersections then produce a reduction in semantic degrees of freedom and thus create, similar to a potential well in physics, local semantic \textit{minima} that can only converge toward stable elements of meaning, which the highest cosine similarities then express. In other words, at these specific categorical intersections, only a reduced number of semantic possibilities can exist. And this corresponds to the human cognitive arguments of advocates for more fixed, invariant, and unified criteria (or at least appearing as such to us) of categorization or similarity mentioned earlier.

In line with our empirical observation, this explanation tends to account for what Bricken et al. (2023) indicate in the realm of synthetic cognition: many neurons appear monosyllabic when viewed with top-activated tokens but are revealed to be polysemous when studied based on tokens with lower activations. Or the fact that Bill et al. (2023), for simplification purposes, limit themselves to interpreting formal neurons only from their few over-activated tokens, noting that including lower activations decreases the explanatory power of their mono-semantic explanations.

\subsection[Human-like Categorical Convergence as a Manifestation of Synthetic Categories at the Interface of Human and Synthetic Cognitions]{"Human-like" Categorical Convergence as a \\ Manifestation of Synthetic Categories at the Interface of Human and Synthetic Cognitions}

We have posited that cases of co-activation of different sub-dimensional categories associated with a neuron generate categorical convergence. But why does this convergence appear to be oriented towards tokens reflecting a relative semantic homogeneity that is isomorphic to human categories of thought, as evidenced by the increase in cosine similarities? One could also imagine that convergence might occur towards tokens whose unity pertains to alien concepts. Below, we discuss the following explanatory hypothesis.

As Thibault (1997) and Roads et al. (2024) mention, commenting on mathematical models of human categorization—such as the "generalized context model" by Nosofsky (1986) and Nosofsky et al. (2022)—the similarity function in these models can be implemented using, for example, a weighted Minkowski distance to formalize the notion of selective attention. The specific attention given to a dimension used for similarity judgment is modeled by the weight assigned to that dimension. The authors state that this selective weighting accounts for a process of contraction or dilation of the cognitive space along that dimension. A high attention weight relative to a dimension stretches the space along that dimension, which has the effect of distancing \textit{stimuli} (in our case, tokens) whose categorical proximity is evaluated on this dimension, thus allowing for greater discrimination of these \textit{stimuli} along this dimension.

By the mathematical construction of the aggregation function, what we just described is precisely the case for tokens with (very) high activation: their activation, for a given neuron, can only be strong insofar as they tend to be co-activated along their sub-dimensional categories (derived from their input vector space) generated by those of their precursor neurons with the strongest connection weights to the neuron involved. Consequently, for these highly activated tokens, the neuron has a greater capacity to discriminate them along the involved sub-dimensional categories.

This mentioned discriminative capacity seems fundamentally to be the function of synthetic neuronal categories, which are \textit{ad hoc} categories (Barsalou, 1995; Glaser et al., 2020; Bove et al., 2022). This is because they are designed to perform fine distinctions between tokens (distinctions gradually constructed by increasingly abstract and subtle synthetic categories) in order to achieve the purpose for which the language model was trained—to predict with very high discriminative accuracy the next token that is precisely relevant (in the case of GPT). However, this predictive precision will only be suitable if it is partially aligned with human semantics because the goal is to generate sentences that are adjusted to this semantics. Therefore, the mentioned discrimination must be, in part (though not entirely, as the efficiency of synthetic categories also derives from their alien concept dimension), convergent with human semantics. This would then be reflected in higher cosine similarities, indicating stronger categorical homogeneity (from the perspective of the semantic universe of human cognition) for tokens with (very) high activations.

To put it another way, more succinctly, at the level of a given neuron, its increased discrimination along certain sub-dimensional categories (derived from its strongly connected precursor neurons) generates a better segmentation and separation of tokens along the different involved sub-dimensions. This then allows for fine categorical intersections of its sub-dimensions, refined abstractions serving the purpose of differentiating predictions of upcoming tokens adapted to human semantic modalities (to which the model was trained).

To conclude this line of inquiry, tokens with (very) high activations can be thought of as being or tending towards the prototypes, in the sense of Rosch (1975) (and also Singh et al., 2020; Vogel et al., 2021), of their respective neurons (the true ultimate prototype being, for a given neuron, the fictitious token that maximizes the aggregation function associated with that neuron). This is because, again by the mathematical construction of the aggregation function, highly activated tokens best satisfy the various sub-dimensional categories (from the input vector space) of their associated neurons. Interpreting things this way supports a strong relationship between categorization and similarity (in line with proponents of indexing categorization on similarity in human cognitive psychology), as we observe that the most strongly activated tokens are associated with increased cosine similarity proximity. But more fundamentally, interpreting core tokens with very high activations as prototypes of the synthetic categories to which they belong allows us to think of these synthetic categories, in their prototypes, as limit cognitive states—phase transition points, to use a physics analogy, at the interface of two sets of constraints that a neural network must reconcile: (i) displaying an output that respects the constraints of human semantics (goal) and (ii) creating artificial categorical segmentations relevant to alien concepts distinct from human semantics but effective in achieving the mentioned purpose (means). The prototypes would then denote an interface subzone category at the boundary (for although the cosine similarity of the tokens involved is higher, the data shows it remains relative) of the informational closure (in Varela's sense) of the alien concepts that are the synthetic concepts of formal neurons.

\section{Conclusion}

In light of our empirical results, which are largely compatible with our hypothesis of categorical convergence—postulating that as the activation levels of successive core tokens increase, the categorical variability of these core tokens decreases—the central explanatory element we have invoked is as follows: the categorical segment constructed by a neuron in a layer n (more precisely by its aggregation function, among other factors) can be decomposed into a vector space of categorical sub-dimensions. These sub-dimensions result from a projection of the input vector space of this neuron, which, by the mathematical construction of its aggregation function, consists of the categorical output dimensions of each neuron in the previous layer $(n-1)$. In other words, to understand a neuron, it must be conceived as being multidimensional, composed of categorical sub-dimensions. Tokens with low activation (single categorical triggered) pertain to these sub-dimensions in a distinct manner, whereas tokens with high activation (co-categorical triggered) involve them jointly, which is reflected in our observation of categorical convergence. We are currently exploring this hypothesis of categorical sub-dimensions within the framework of a "genetic" study aimed at explaining the categorical abstraction produced by artificial neurons in terms of the reconstruction of the categorical segmentations of their most influential precursor neurons (i.e., those with the strongest neural connections).

\subsection*{Acknowledgements}

Michael Pichat thanks Christian Ganem (Chrysippe-R\&D) for his valuable advice on innovative readings in the field of AI, Pierre Laniray (Université Paris Dauphine-PSL) for his encouragement to study issues of explainability in AI, Stéphane Fadda (Université Sorbonne) for his stimulating advice in the realm of AI, and Alexander Krainov (Yandex) for the challenging discussions we had regarding the relevance of studying the psychology of AI.

\subsection*{Author Contributions}

Michael Pichat conceptualized and designed the study and served as the scientific lead. Enola Campoli contributed to various operational aspects of the study. William Pogrund handled data preprocessing and statistical processing. Michael Veillet-Guillem managed the software infrastructure part of the study. Jourdan Wilson participated in prompt engineering activities, translated the text into English, and formatted the published text. Anton Melkoezrov was involved in prompt engineering activities and formatted the tables and diagrams. Samuel Demarchi provided advice on statistical studies. Judicael Poumay formated the initial data and advised about conceptual issues. Armanush Gasparian and Paloma Pichat made the realization of this study possible and scaffolded it at different levels.

\section*{Appendices}
\subsection*{Normality of Regression Residuals of Cosine Similarity Values of Pairs of Successive Core-Tokens as a Function of Activation of the First Token of Each Pair (Witness Neurons)}

\begin{figure}[H]
  \centering
  \includegraphics[width=1\textwidth]{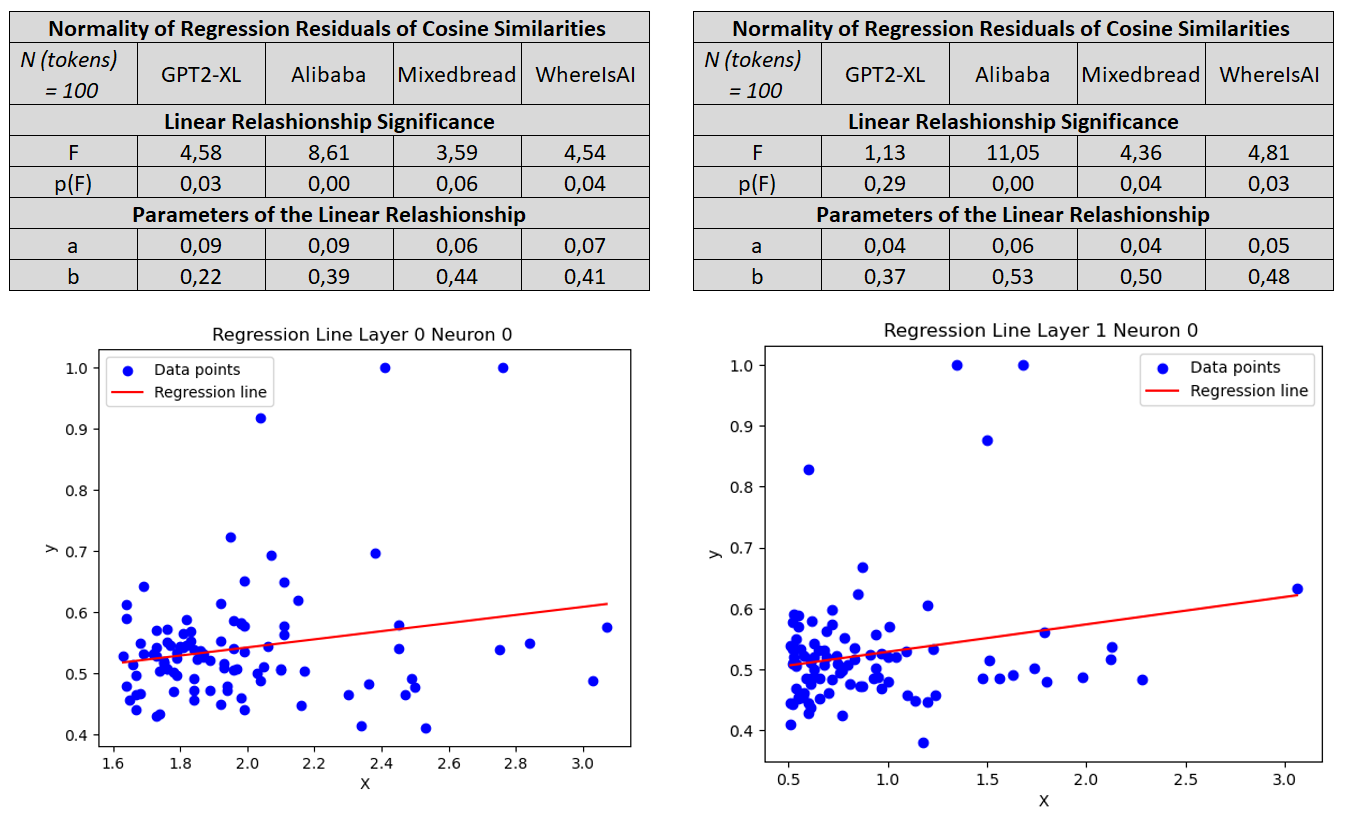}
\end{figure}

\begin{figure}[H]
  \centering
  \includegraphics[width=1\textwidth]{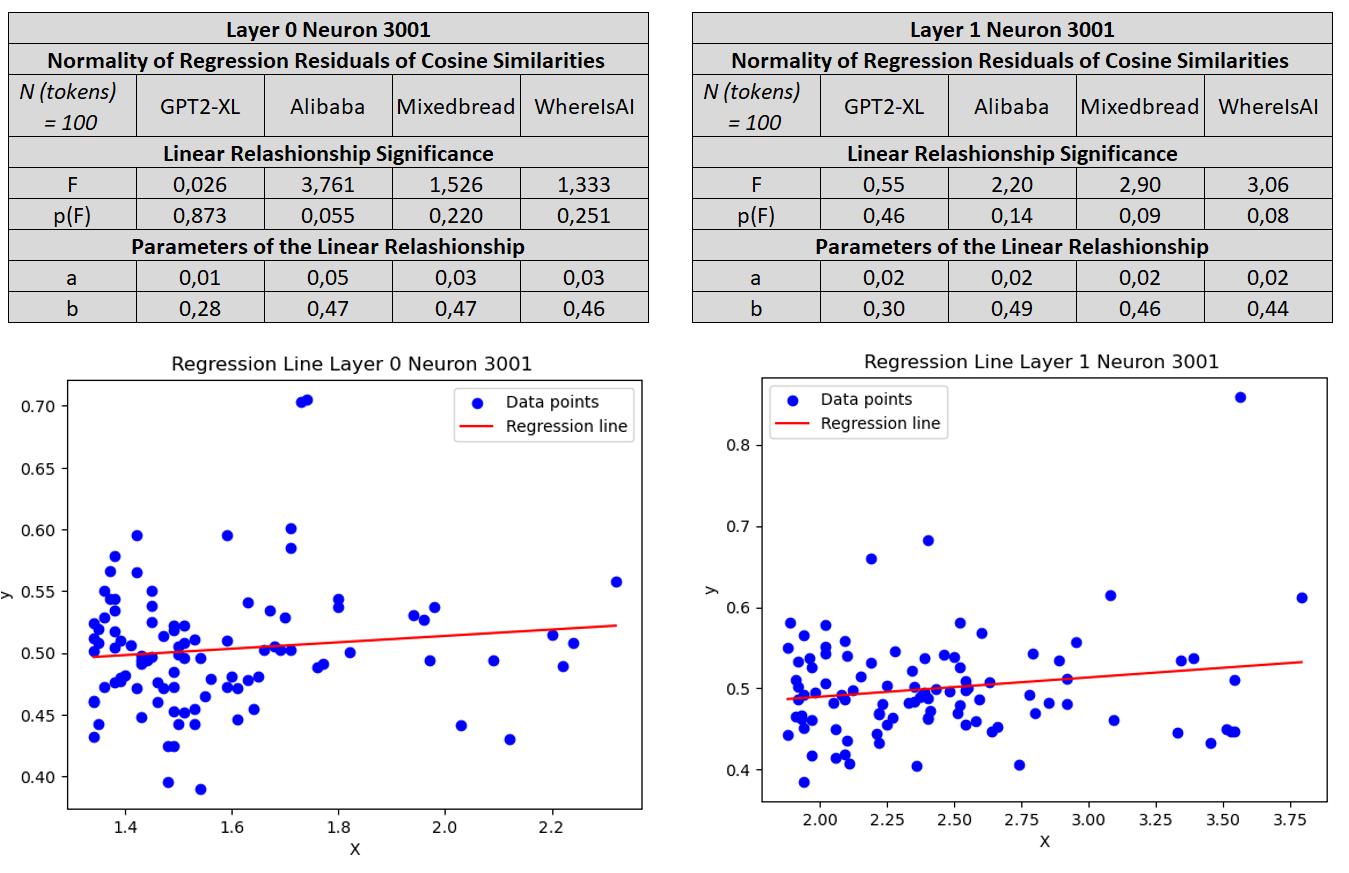}
\end{figure}

\end{document}